\documentclass[a4paper,11pt]{article}
\pdfoutput=1 

\usepackage{jheppub} 

\usepackage[T1]{fontenc} 
\usepackage{float}
\usepackage{ragged2e}
\usepackage{comment}
\usepackage{multicol}
\usepackage[usenames,dvipsnames]{xcolor}

\title{\justify \textbf{Exploring the structure of hadronic showers and the hadronic energy reconstruction with highly granular calorimeters} \small{(LP2021), 10-14 January 2022.}}

\author[\dagger, 1]{H. García Cabrera\note{Speaker}. On behalf of the CALICE Collaboration.}

\affiliation[\dagger]{Centro de Investigaciones Energéticas Medioambientales y Tecnológicas (CIEMAT). Madrid, Spain.}

\emailAdd{hector.garcia2@ciemat.es}

\abstract{Prototypes of electromagnetic and hadronic imaging calorimeters developed and operated by the CALICE collaboration provide an unprecedented wealth of highly granular data of hadronic showers for a variety of active sensor elements and different absorber materials. In this presentation, we discuss detailed measurements of the spatial and the time structure of hadronic showers to characterise the different stages of hadronic cascades in the calorimeters, which are then confronted with GEANT4-based simulations using different hadronic physics models. These studies also extend to the two different absorber materials, steel and tungsten, used in the prototypes. The high granularity of the detectors is exploited in the reconstruction of hadronic energy, both in individual detectors and combined electromagnetic and hadronic systems, making use of software compensation and semi-digital energy reconstruction. The results include new simulation studies that predict the reliable operation of granular calorimeters. Further we show how granularity and the application of multivariate analysis algorithms enable the separation of close-by particles. We will report on the performance of these reconstruction techniques for different electromagnetic and hadronic calorimeters, with silicon, scintillator and gaseous active elements. Granular calorimeters are also an ideal testing ground for the application of machine learning techniques. We will outline how these techniques are applied to CALICE data and in the CALICE simulation framework.}

\begin{document} 
\maketitle 

\section{Introduction}
The CALICE collaboration is an R\&D group of around 280 physicist and engineers from around the world, working together to develop new high granularity detectors for high energy e+e- experiments. These detectors are being developed for possible future experiments like the ILC \cite{behnke2013international}, CLIC \cite{linssen2012physics}, FCCee \cite{FCCee}, CEPC \cite{thecepcstudygroup2018cepc} or smaller projects like LUXE \cite{LUXE2021}. One of the main objectives in these e+e- experiments is to perform high precision physics measurements and also model independent Higgs analysis to study its properties and possibly extend the analysis into the discovery of new particles beyond the scope of the LHC \cite{baer2013international}. To fully exploit the potential of future experiments, a precise reconstruction of all final states is necessary. A main technological challenge is to achieve a jet energy resolution in the order of 3\% - 4\% over a wide range of energies allowing the identification of the Z or W bosons in hadronic decays. The approach selected to reach this resolution is to implement Particle Flow Algorithms (PFA) where the measurement of different type of particles is performed in the most appropriate sub-detector for the energy and momentum measurements. Charged particles are measured in the tracker, for photon energies the electromagnetic calorimeter and in the case of neutral hadrons they are measured in the hadronic calorimeter. Since charged particles and photons carry about 90\% of the energy, PF intrinsically increases the jet energy resolution. To efficiently apply PF each energy contribution need to be separated and identified for which high transverse and longitudinal granularity should be provided by the calorimeters. Since 2005 different calorimeter technologies are being studied by the CALICE collaboration with the scope of PF in their designs. \\

\begin{figure}[b]
    \centering
    \includegraphics[width=0.8\textwidth]{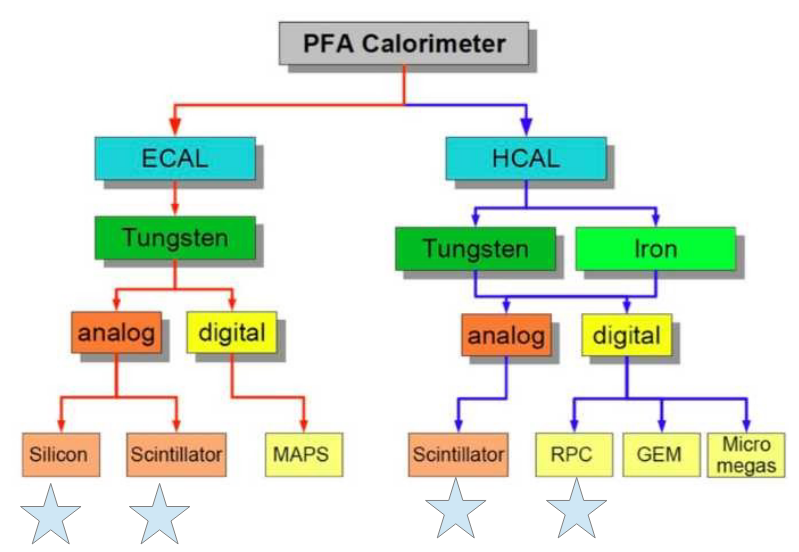}
    \caption{High granularity technologies studied in the CALICE collaboration. The stars represent the projects with a technological prototype currently working. }   
    \label{fig:Technologies}
\end{figure}

Figure \ref{fig:Technologies} is a schema of the projects hosted by CALICE.

\section{Monte Carlo simulations}

To build reliable PFA a detailed study of the calorimeter response to particle interactions is necessary. This implies the precise simulation and reconstruction of the interaction of neutral and charged hadrons using the subsequent particle cascade. The simulations were carried out with the Mokka \cite{MoradeFreitas:2002kj} framework which provides the geometry interface to Geant4 \cite{Allison:2016lfl} in which several interaction models are combined into the simulation models, also called physics lists. \\

\begin{figure}[h]
    \centering
    \includegraphics[width=\textwidth]{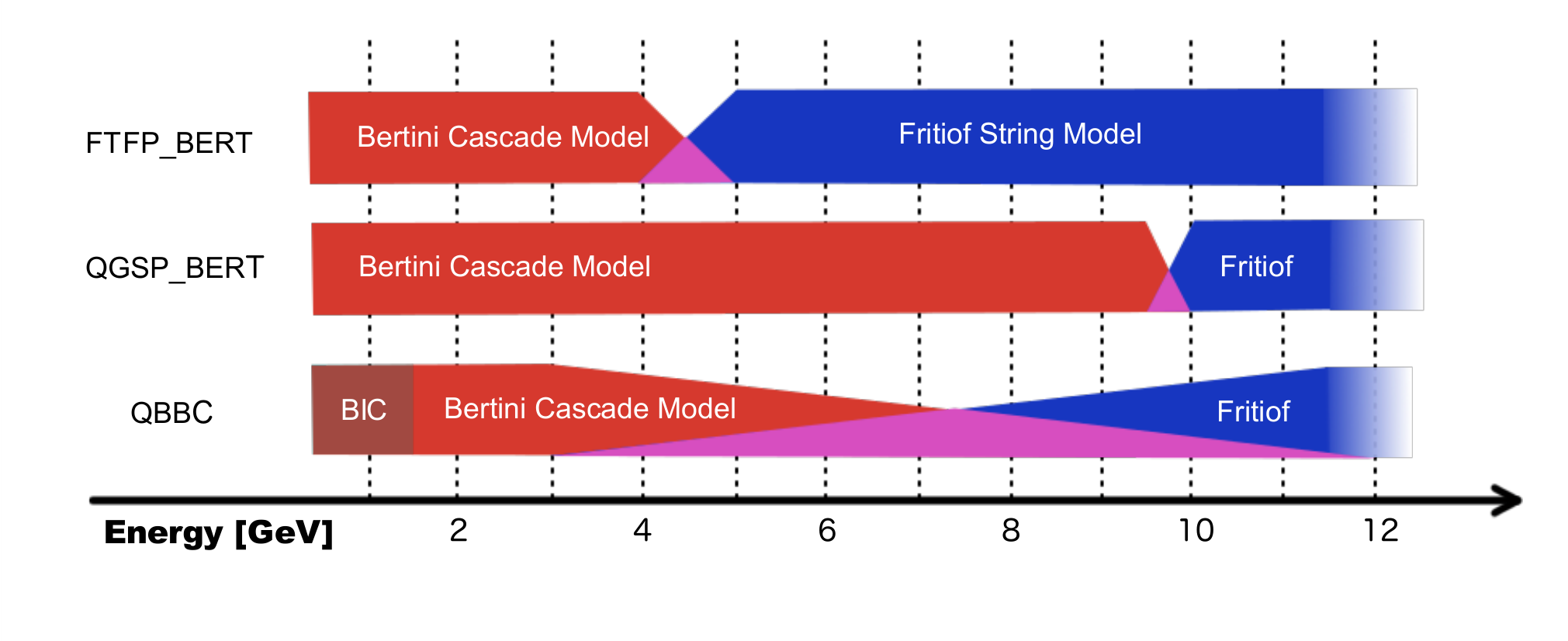}
    \caption{Hadronic physics lists studied, they are a combination of the Bertini Cascade model, the Fritiof String Model and the Binary Cascade model in different energy ranges and interpolations. }
    \label{fig:MCModels}
\end{figure}

Figure \ref{fig:MCModels} shows the simulation models that have been studied together with the electromagnetic shower models using different prototypes. \\

\noindent Analysis performed with the Digital Hadronic Calorimeter (DHCAL) have shown a strong dependence of the response and the energy resolution on the electromagnetic physics lists for positron and pion showers being the QGSP\_BERT\_EMZ physics list the one with the overall best agreement with the data \cite{2019}. Also the Silicon Tungsten Electromagnetic Calorimeter (SiWECAL) \cite{Kawagoe:2019dzh} simulations reproduce the behaviour of the hadronic showers within a 10\% discrepancy without showing a clear preference for one simulation model \cite{2019Tracks}. \\

\section{Particle Flow Algorithms}

\begin{figure}
    \centering
    \includegraphics[width=0.47\textwidth]{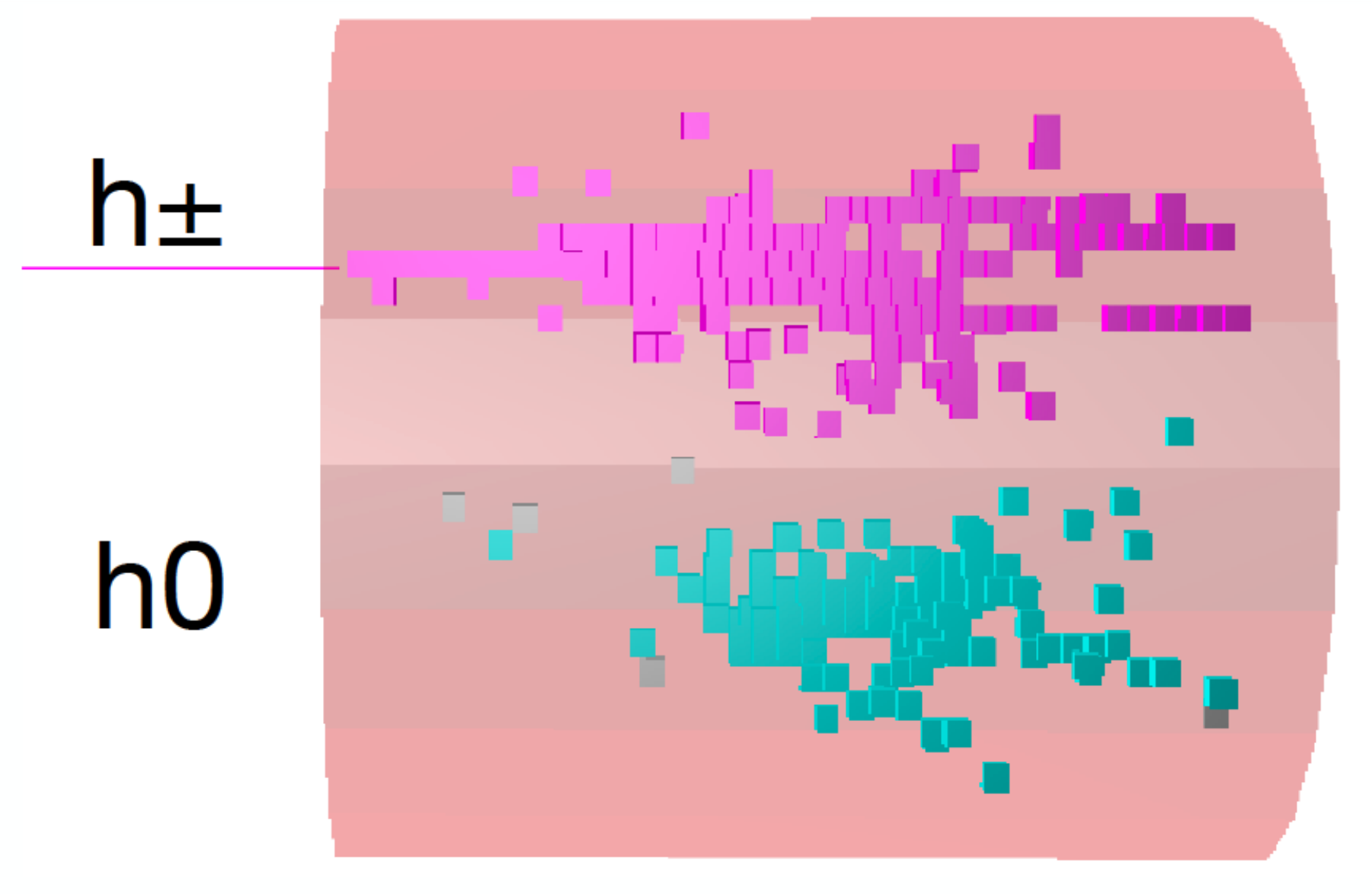}
    \includegraphics[width=0.49\textwidth]{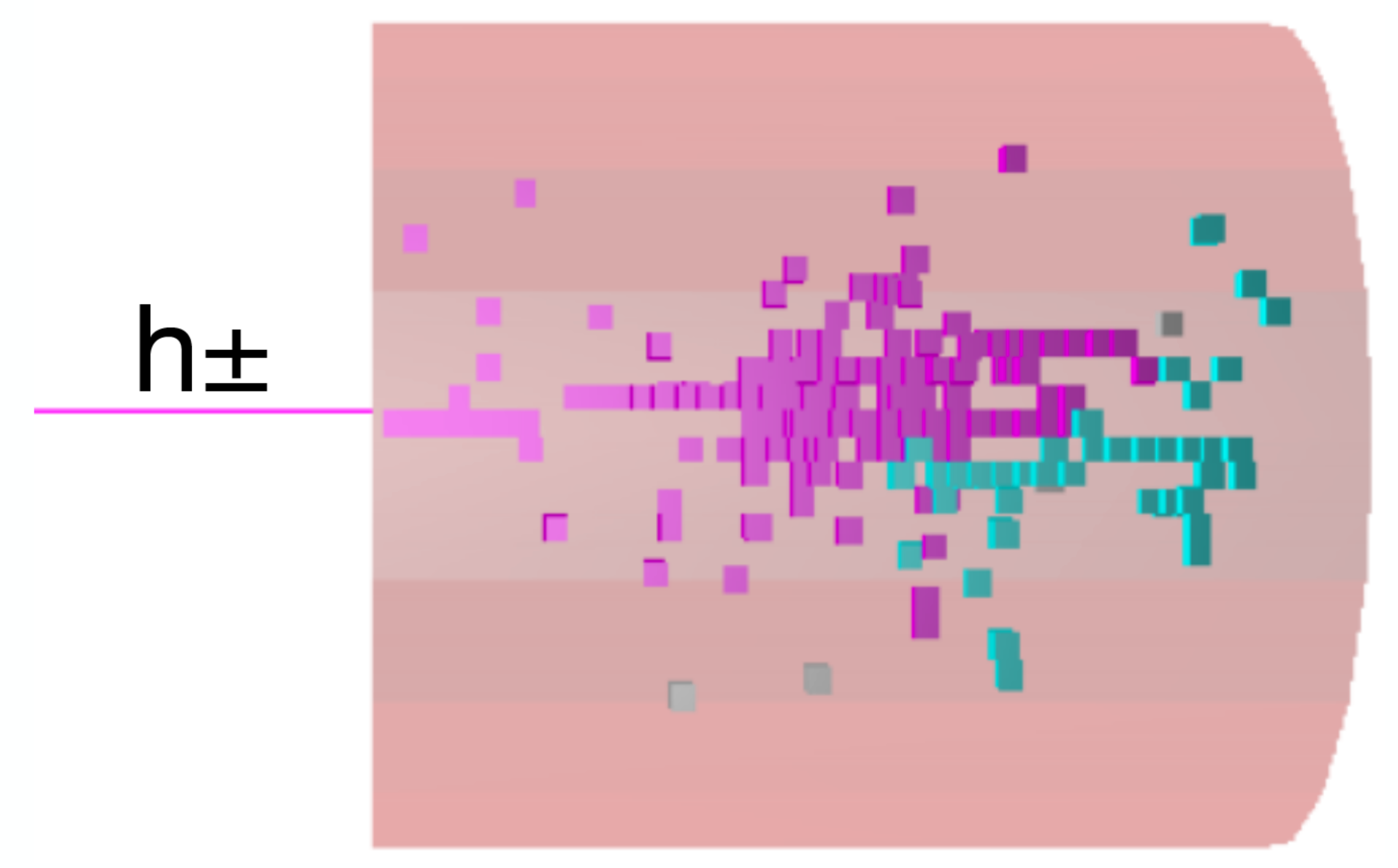}
    \caption{An example of a good separation of a charged from a neutral hadron (left) and of an excess of neutral energy that degrades the energy resolution (right).}
    \label{fig:Separation}
\end{figure}

\noindent Ideally in PF only the energy of neutral particles is measured in the calorimeters while the charged particles are reconstructed in the tracker where the resolution is much better. Then the capability of the PFA to recover neutral hadrons in the vicinity of a charged hadron is of crucial importance because miss-assignments of energy would degrade the energy resolution. \\

as shown in Figure \ref{fig:Separation}.\\ 

\begin{figure}[h]
    \centering
    \begin{multicols}{2}
    \includegraphics[width=0.37\textwidth]{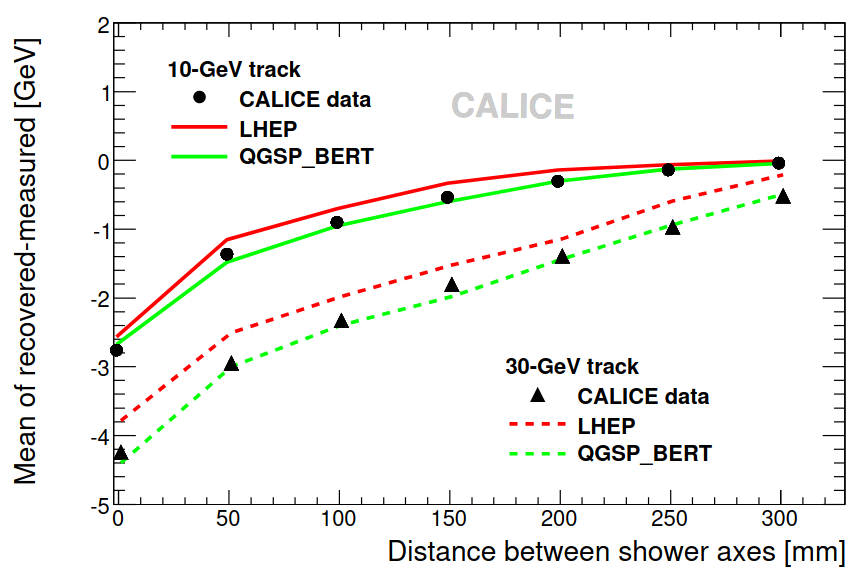}
    \includegraphics[width=0.3\textwidth]{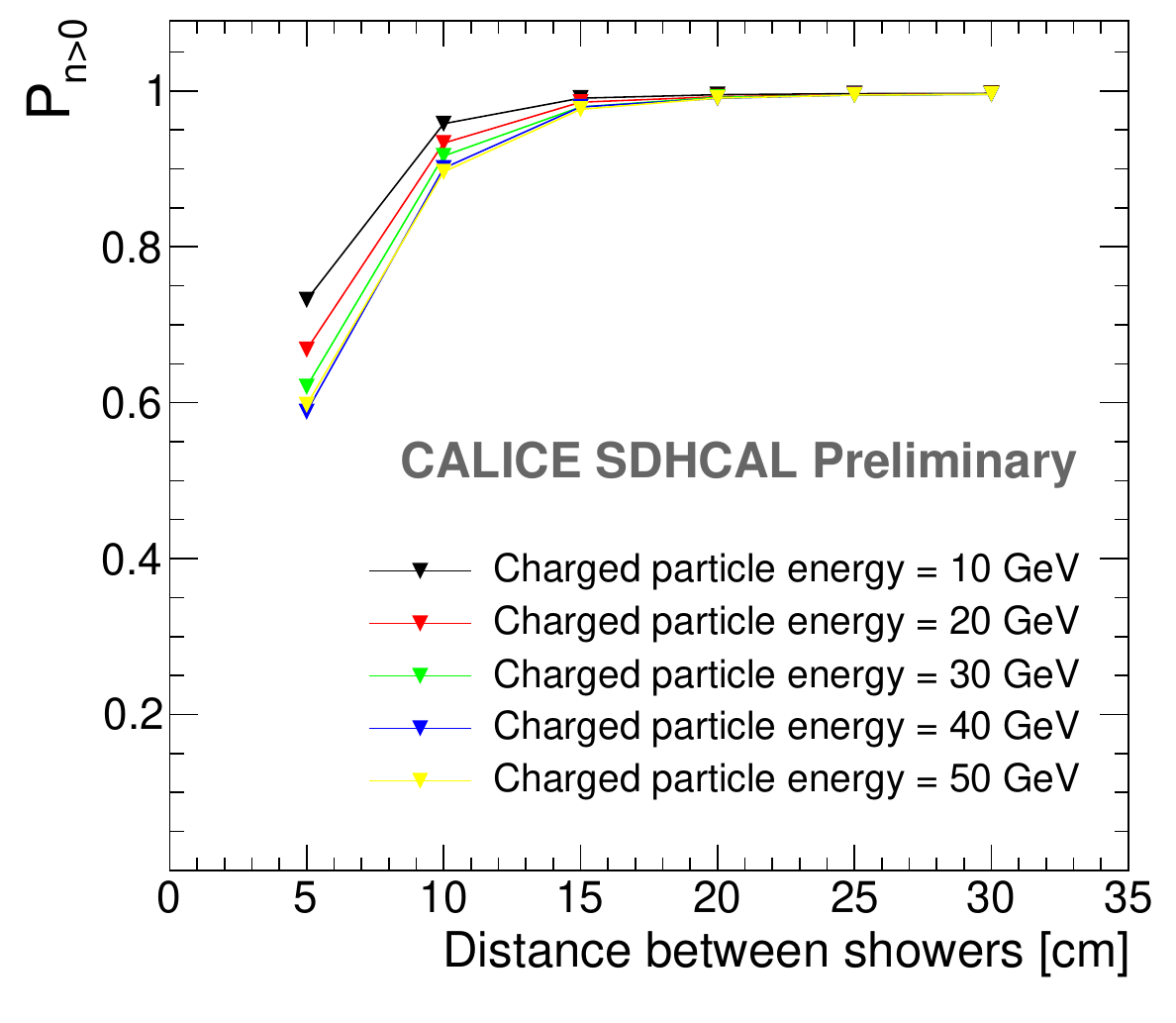}
    \end{multicols}
    \caption{Results of the analysis of the PFAs. A comparison of the difference of the mean recovered energy and the measured energy versus the distance of the hadron axes with AHCAL data and simulations (left). Efficiency of recovering a neutral hadron as a function of the distance between the shower axes (right).}
    \label{fig:PFAResults}
\end{figure}

\noindent One of the most developed PFA in the context of the ILC is Pandora \cite{2009}, which is the one used by default in the reconstruction chain. The performance of this algorithm has been tested with the large Analog Hadronic Calorimeter (AHCAL) \cite{2019AHCal} physics prototype both with data of pions from 10 - 30 GeV and simulations of the ILC detector. This validation would provide further evidence that the particle flow reconstruction obtained in a simulation of the ILC detector concept is realistic. To reproduce this study in the scenario with the AHCAL two hadron events have to be overlayed and then the track of one hadron before the shower start is removed to emulate the neutral hadron. Later such event is mapped into the ILD geometry. Several separation distances between shower axes were tested. The left image in Figure \ref{fig:PFAResults} shows that at small distances between particles, where the shower overlap is considerable, the mean energy of the neutral hadron recovered is typically lower than the corresponding energy measured in the calorimeter prototype. Nevertheless, simulations and data are in good agreement demonstrating that the extrapolation to the complete detector is reliable, with the QGSP\_BERT simulation model giving the better description of the data \cite{thecalicecollaboration2011tests}. \\

\noindent Another PFA is the Arbor \cite{ruan2014arbor} approach, studied with the Semi-Digital Hadronic Calorimeter (SDHCAL) \cite{baulieu2015construction} technological prototype with data of pions from 10 - 80 GeV. Test of single particle and multi particle separation were performed. This algorithm is based on the idea that the hadronic shower development follows a tree-like topology, a principle close to the underlying physics. In the single particle analysis the hit clustering efficiency shows a constant value over 96\% in the studied range. However, since the number of hits increases with the energy the number of missed hits also increases and the number of reconstructed particles is larger due to shower splitting. For the multi-particle separation, similar to the Pandora analysis, pion events are overlayed testing various separation distances. As expected, in the plot in right of Figure \ref{fig:PFAResults} the confusion grows larger at lower distances giving an efficiency of recovering a neutral hadron higher than 90\% down to 10 cm. At small distances the reconstruction of the neutral hadron has a binary behaviour, either is very well reconstructed or merged into the charged hadron \cite{CALICE:2015njp}. \\

\section{Hadronic shower structure}

\begin{figure}[b]
    \centering
    \includegraphics[width=0.3\textwidth]{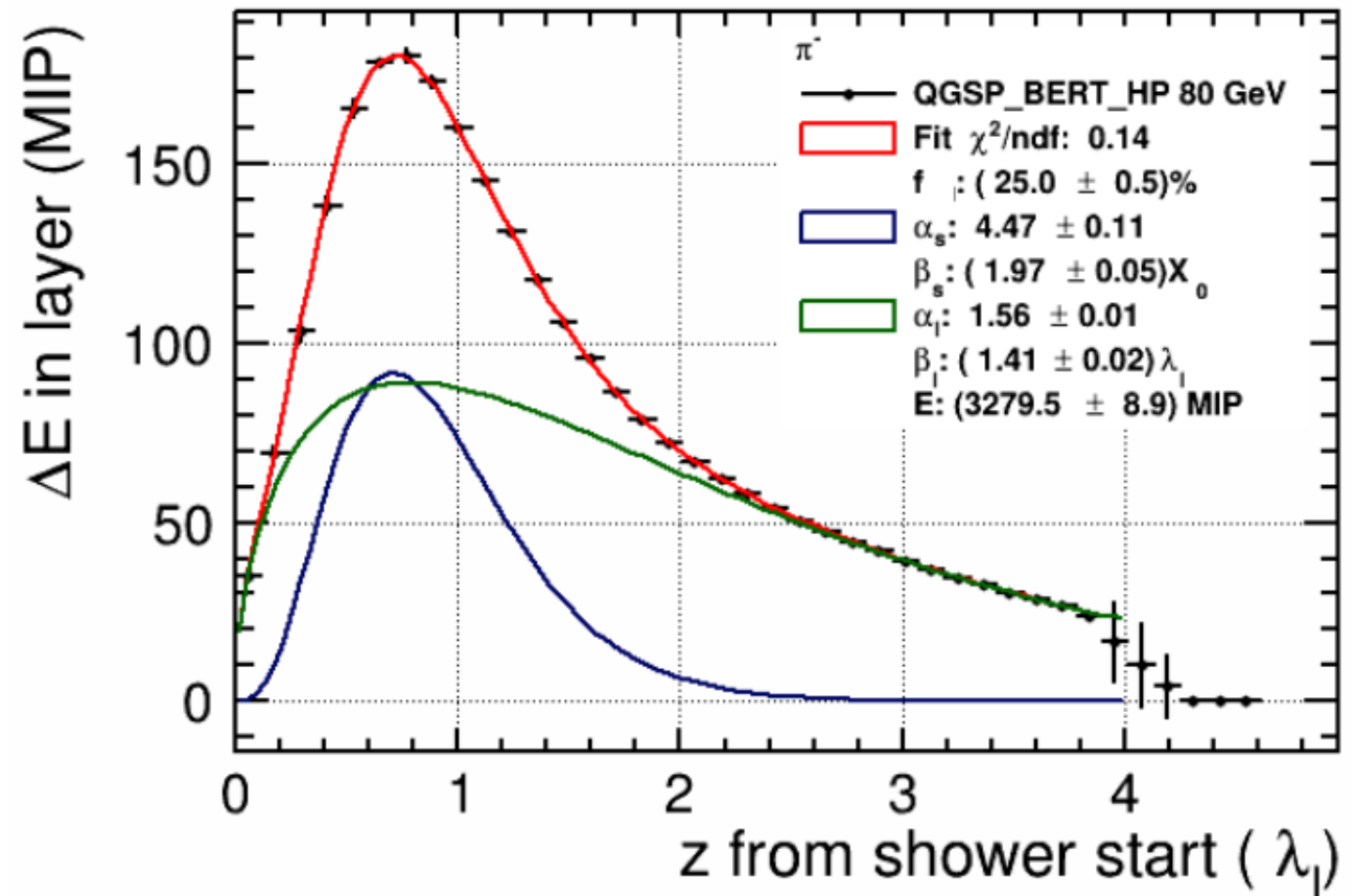}
    \hspace{70pt}
    \includegraphics[width=0.3\textwidth]{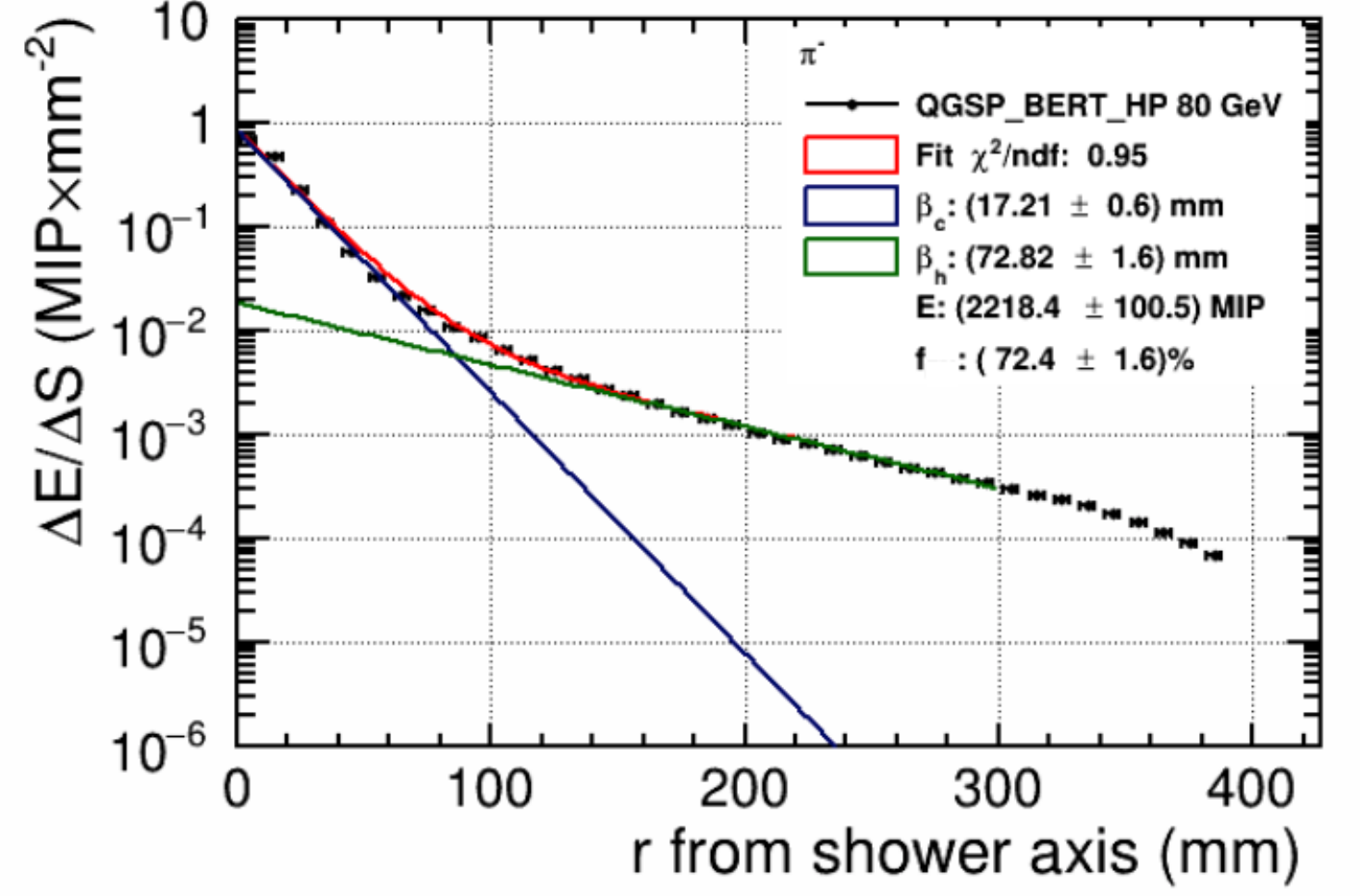}
    \caption{Longitudinal (left) and radial (right) shower development of simulated 80 GeV pions with fitted parametrizations (1) and (2).}
    \label{fig:HadronicStructure}
\end{figure}

\noindent The high spatial granularity and longitudinal segmentation of the prototypes offer unique possibilities to study the shape of hadronic showers. For example, by investigating the longitudinal and radial energy density, a shower can be separated into the electromagnetic core and the halo part. The longitudinal parametrization is composed of a sum of incomplete gamma functions: \\

\begin{equation}
    \Delta E(Z) = E \cdot \left( \textcolor{blue}{\frac{f}{\Gamma (\alpha_s)} \left( \frac{Z[X_0]}{\beta_s} \right)^{\alpha_s - 1} \cdot \frac{e^{\frac{Z[X_0]}{\beta_s}}}{\beta_s}} + \textcolor{OliveGreen}{\frac{1 - f}{\Gamma (\alpha_l)} \left( \frac{Z[\lambda_l]}{\beta_l} \right)^{\alpha_l - 1} \cdot \frac{e^{\frac{-Z[\lambda_l]}{\beta_l}}}{\beta_l}}\right)
\end{equation}

\noindent where $\alpha_s$ and $\beta_s$ are shape parameters of the core components, $\alpha_l$ and $\beta_l$ are the shape parameters of the halo component, f is the relative weight of core to halo component, Z[$\lambda_l$] is the depth in the calorimeter in terms of the nuclear interaction length and Z[$X_0$] is the depth in the calorimeter in terms of the radiation length. For the radial profile the parametrization is: \\

\begin{equation}
\frac{\Delta E}{\Delta S} (r) = \frac{E}{2\pi} \left( \textcolor{blue}{f \cdot \frac{e^{\frac{-r}{\beta_c}}}{\beta^2_c}} + \textcolor{OliveGreen}{(1 - f)\cdot \frac{e^{\frac{-r}{\beta_h}}}{\beta^2_h}} \right)
\end{equation}

\noindent with r as the radius from the shower center and $\beta_c$ and $\beta_h$ as the shape parameters of the core and halo part. The parameters can be computed from a fit to the simulated shower as has been done with the AHCAL \cite{emberger2021analysis}, shown in Figure \ref{fig:HadronicStructure}. \\

\section{Secondary Tracks}

\noindent Another ability thanks to the high granularity of the prototypes is to compute secondary tracks within an hadronic shower. Each technology has developed its particular algorithm, being a backwards iterative finding algorithm \cite{2019Tracks}, like the SiWECAL, or Hough transformed based, for the case of the hadronic calorimeters \cite{2017} \cite{calicecollaboration2013track}. The identification of secondary tracks is a quite powerful tool. For example, in the case of the SDHCAL, low energy particles that stop inside the calorimeter may have hits with high threshold values. These high threshold hits within a secondary track may bias the energy estimation. Therefore giving the same weight for all the hits in a track has improvements into the energy reconstruction. \\

\section{Multi Variate Analysis}

\begin{figure}[h]
    \centering
    \includegraphics[width=0.3\linewidth]{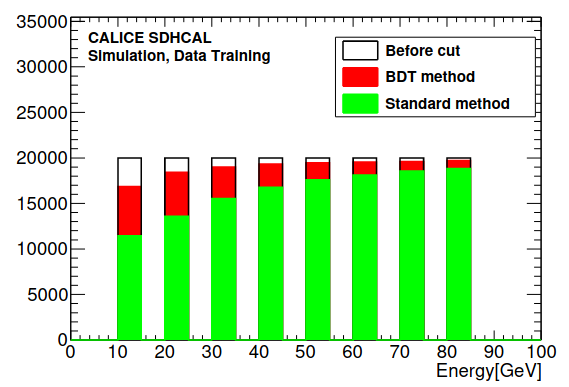}
    \hspace{70pt}
    \includegraphics[width=0.32\linewidth]{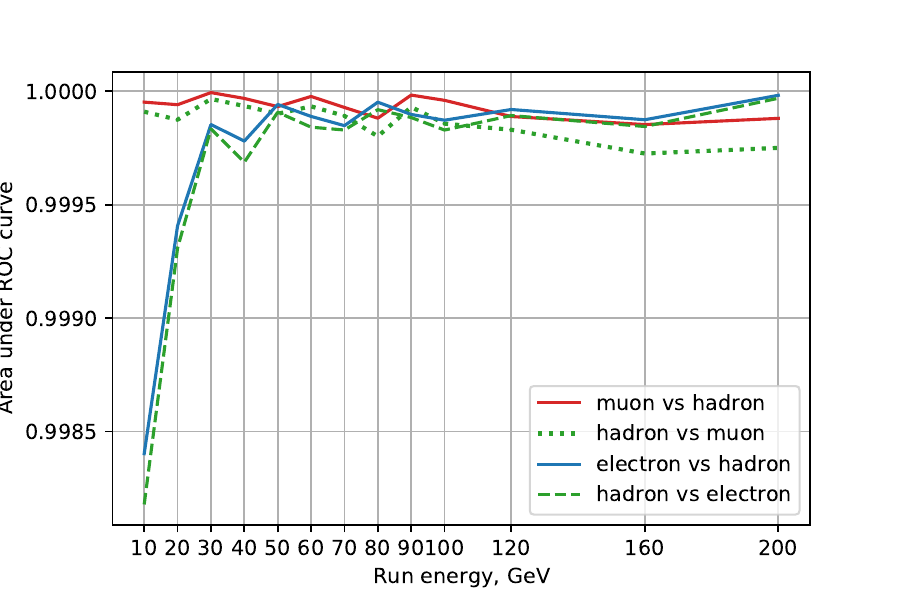}
    \caption{Improvement in the selection efficiency of the BDT method over the SDHCAL standard method (left). Particle identification performance of the three classifiers (electron, hadron and muon) in terms of the area under the ROC curve for the AHCAL simulated data (right).}
    \label{fig:BDT}
\end{figure}

\noindent Efficient identification of particles is a key ingredient to successful PFAs to avoid double counting of particles. Thanks to the high granularity of the detectors it is possible to use topological variables to identify the particle crossing the calorimeters. However, for a proper particle identification several observables are needed, giving rise to a multi variate approach. These methods make use of machine learning techniques where the inputs are the set of observables that produces an output capable of classifying an event as signal or background. The method developed by the SDHCAL \cite{2020} uses a Boosted Decision Tree (BDT) creating two classifiers: one to separate showers from muons and another to separate the hadronic from the electromagnetic showers. A cut is applied to the output of the BDT to separate the particle type, then the two classifiers are applied in order, first the pion - muon classifier and then the pion - electron classifier. Results shown in the left plot of Figure \ref{fig:BDT} have a significant statistical gain with respect to the standard method and with a stable performance over a wide range of energies. Another BDT method is currently being developed with simulations of the full AHCAL prototype \cite{emberger2021analysis}, the right of Figure \ref{fig:BDT} shows the early results of the separating power over the simulated energy range with an excellent performance. The measure of separation power is the area under the receiver operating characteristic (ROC\footnote{A ROC curve describes the performance of a classification model at all classification thresholds by showing the ratio of false positive rate to true positive rate. The integral of this curve measures the identification power of the model, where 0 characterizes a 100\% wrong model and 1 characterizes a 100\% correct model.}) curve \cite{emberger2021analysis}.\\ 
 
\section{Software compensation}

\begin{figure}[h]
    \centering
    \includegraphics[width=0.3\linewidth]{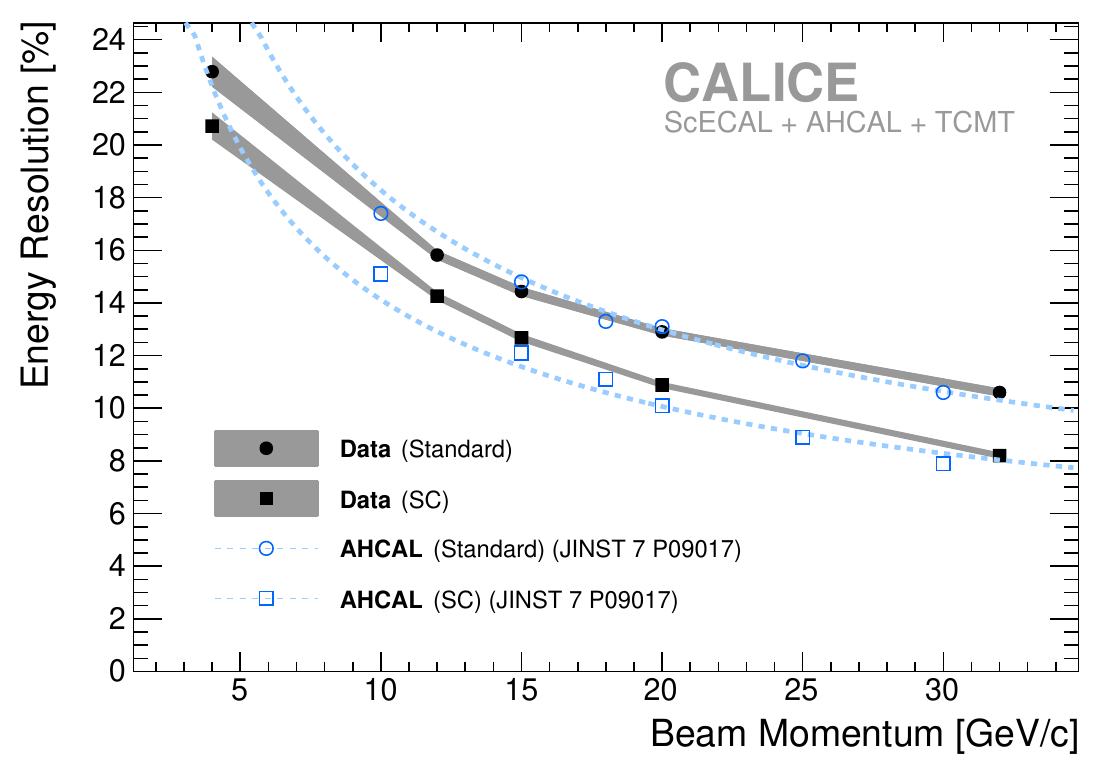}
    \hspace{70pt}
    \includegraphics[width=0.3\linewidth]{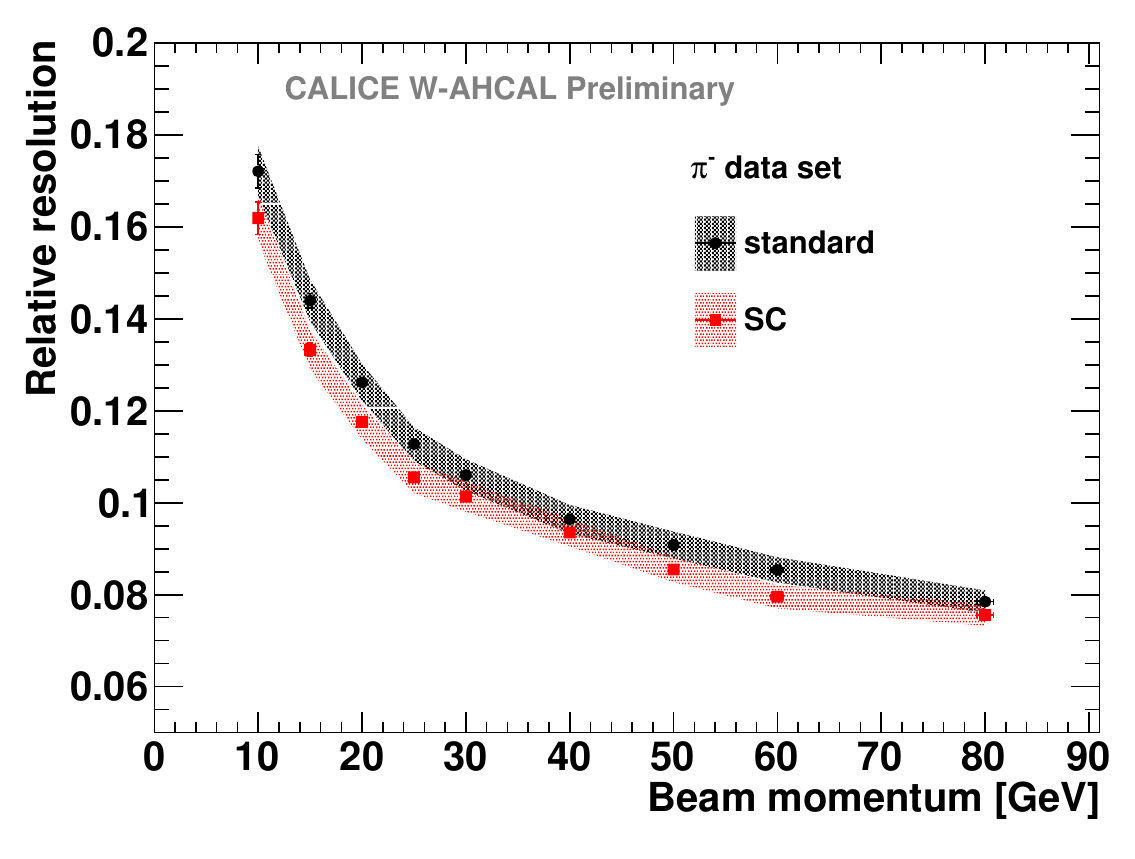}
    \caption{Energy resolution improvement from applying software compensation using the technique of hit weighting with the Fe absorber (left) or linearisation in the W absorber (right). }
    \label{fig:SC}
\end{figure}

\noindent In addition to the standard energy reconstruction methods, it is possible to apply software compensation techniques, in the case of non-compensating calorimeters like the AHCAL, to correct the different response to an hadronic shower than the response to an electromagnetic shower of the same initial energy. These methods consist of correcting the shower energy with a set of linearisation coefficients tuned using a set of negative pion samples or weighting differently the low energy hits dominated by the hadronic component from the high energy hits dominated by the electromagnetic components. The energy resolutions in Figure \ref{fig:SC} improve from the standard method by a 10\% up to a 20\% for the AHCAL with Fe absorber \cite{calicecollaboration2018hadronic} while in the W case this improvement is not so significant due to a better level of compensation \cite{CALICE:2018otk}.\\

\section{Timing}

\begin{figure}[h]
    \centering
    \begin{multicols}{2}
    \includegraphics[width=0.32\textwidth]{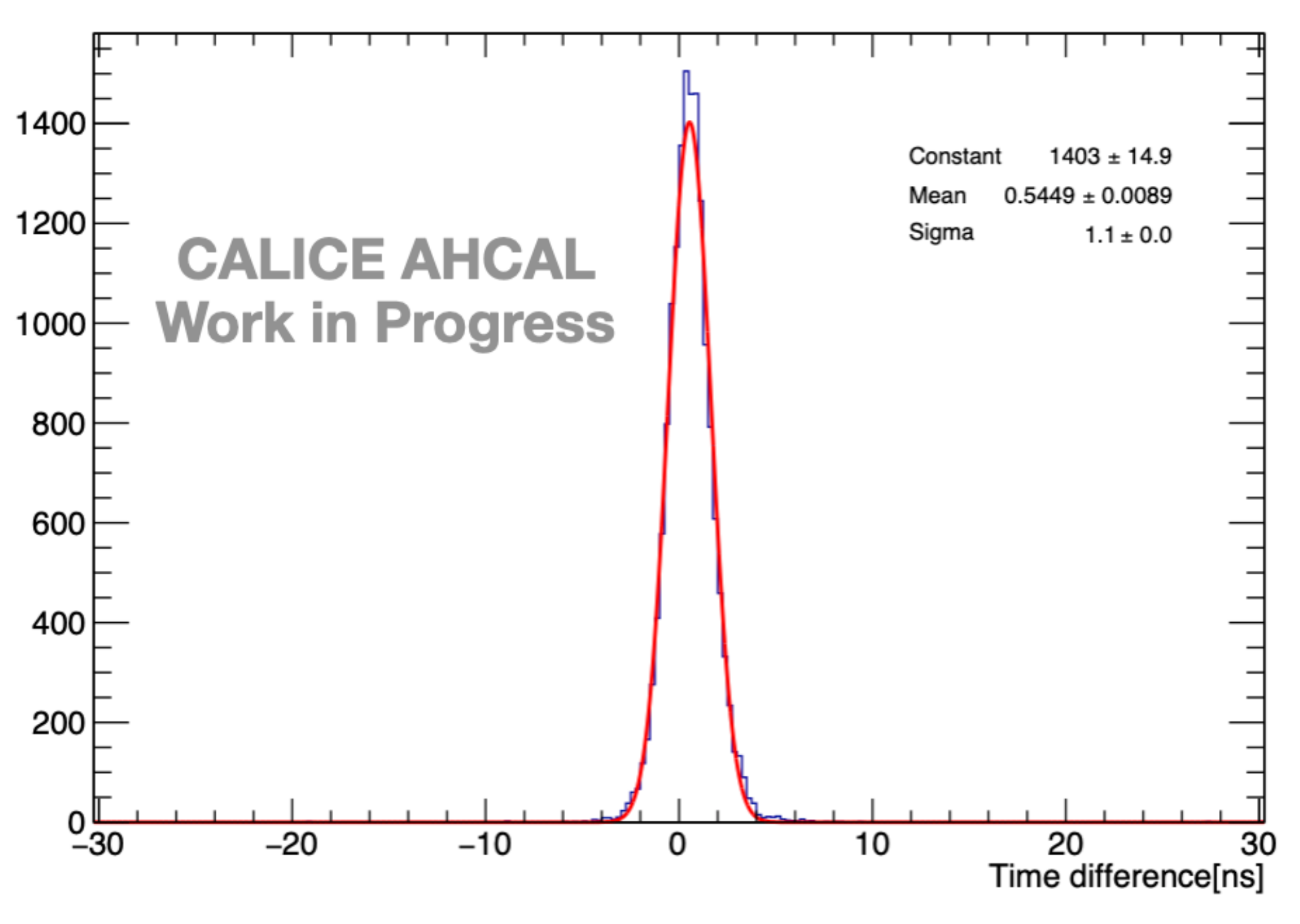}
    \vfill \null
    \columnbreak
    \includegraphics[width=0.25\textwidth]{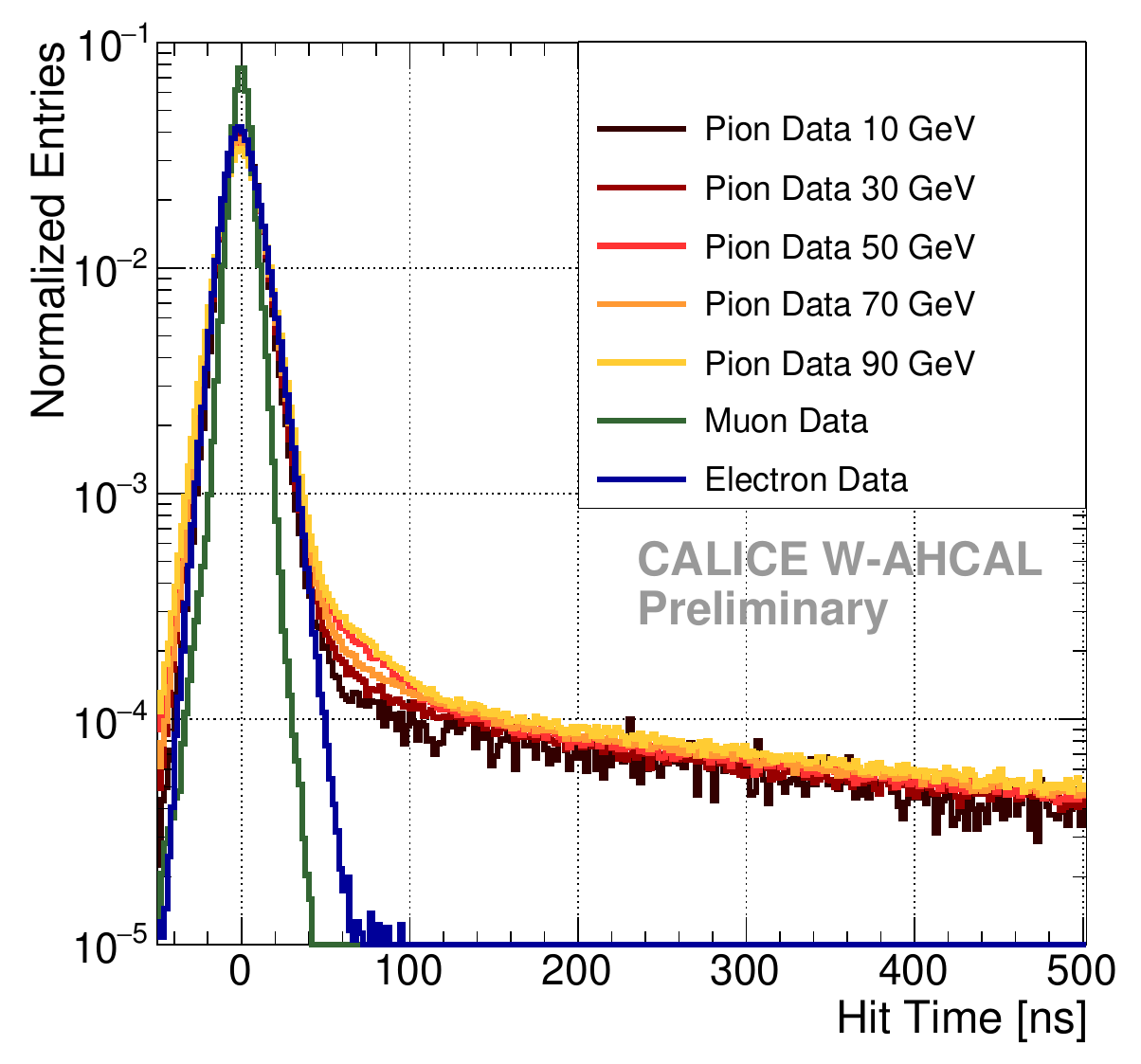}
    \end{multicols}
    \caption{Hit time resolution (left) and hadronic time structure (right) with the AHCAL technological prototype.}
    \label{fig:Timing}
\end{figure}

\noindent It is in the scope of many technologies to include precise timing information to study the complex time structure of hadronic showers. Early studies \cite{CALICE:2019ipm} have been performed with the AHCAL where the time structure of pions have been computed with a hit time resolution of ~1 ns as shown in Figure \ref{fig:Timing}. The tail in the pion spectrum is due to the presence of low energetic neutrons. However, this resolution is not for precision measurements, better resolution is expected in the future in order of a few hundred of ps. \\

\section{Summary}
 
 \noindent This article summarizes the performed and on-going studies with the different detector prototypes developed by CALICE. Results show that high granularity is one of the key components to reach the unprecedented jet energy resolution needed for future e+e- experiments. Also their imaging capabilities provide an excellent opportunity to study the structure of hadronic showers and calorimeter based particle identification that allow for validation of the Geant4 models, improvements to the PFA performance and corrections to the energy reconstructions improving the resolution of hadrons. Finally, the timing measurements show promising results for the upcoming developments. \\
 
\newpage

\bibliographystyle{unsrt}
\bibliography{main.bib}
 
\end{document}